\documentclass[prl,reprint]{revtex4-1}

\usepackage{graphicx}
\usepackage{bm}
\usepackage{hyperref}
\usepackage{color}

\begin{document}

\title{Spin and charge distribution symmetry dependence of stripe phases in two-dimensional electron systems confined to wide quantum wells}

\author{Yang Liu}
\affiliation{Department of Electrical Engineering,
Princeton University, Princeton, New Jersey 08544}
\author{D. Kamburov}
\affiliation{Department of Electrical Engineering,
Princeton University, Princeton, New Jersey 08544}
\author{M.\ Shayegan}
\affiliation{Department of Electrical Engineering,
Princeton University, Princeton, New Jersey 08544}
\author{L.N.\ Pfeiffer}
\affiliation{Department of Electrical Engineering,
Princeton University, Princeton, New Jersey 08544}
\author{K.W.\ West}
\affiliation{Department of Electrical Engineering,
Princeton University, Princeton, New Jersey 08544}
\author{K.W.\ Baldwin}
\affiliation{Department of Electrical Engineering,
Princeton University, Princeton, New Jersey 08544}

\date{\today}

\begin{abstract}

  Measurements in clean two-dimensional electron systems confined to
  wide GaAs quantum wells in which two electric subbands are occupied
  reveal an unexpected rotation of the orientation of the stripe phase
  observed at a half-filled Landau level. Remarkably, the
  reorientation is sensitive to the spin of the half-filled Landau
  level and the symmetry of the charge distribution in the quantum
  well.
\end{abstract}


\maketitle

A low-disorder two-dimensional electron system (2DES) subjected to a
strong perpendicular magnetic field ($B$) displays a variety of novel
quantum phases. At high $B$, when the Fermi energy ($E_F$) resides in
the lowest ($N=0$ and 1) Landau levels (LLs), electrons typically
condense into incompressible liquid states and exhibit the fractional
quantum Hall effect \cite{Tsui.PRL.1982}. At lower $B$, when $E_F$
lies in the higher LLs ($N\ge 2$), phases with non-uniform density are
predicted to be the ground states. More specifically, when a
spin-split $N\ge 2$ LL is half filled, the 2DES breaks the rotational
symmetry by forming a unidirectional charge density wave, the
so-called stripe phase \cite{Koulakov.PRL.1996,
  Moessner.PRB.1996}. Experimentally, strong anisotropy is seen in
in-plane transport coefficients at LL filling factors $\nu=9/2$, 11/2,
13/2, and 15/2: the longitudinal resistance commonly vanishes along
the [110] crystal direction along which the stripes form ("easy"
axis), but exhibits a strong peak along the [1\=10] direction ("hard"
axis) \cite{Lilly.PRL.1999, Du.SSC.1999}. It is believed that a native
symmetry-breaking field, which is still unidentified after more than a
decade of research, is responsible for orienting the stripe phases
along [110] \cite{Cooper.SSC.2001, Willett.PRL.2001}. The associated
anisotropy energy is estimated to be a few mK per electron
\cite{Jungwirth.PRB.1999, Stanescu.PRL.2000} from the fact that an
in-plane magnetic field of $\sim$ 1 T can overcome this energy and
re-align the stripes to be perpendicular to the field direction
\cite{Pan.PRL.1999.Stripe, Lilly.PRL.83.1999, Note1}. Even without any
external symmetry-breaking field, the stripes are known to rotate from
the "normal" ([110]) direction to the "abnormal" ([1\=10]) direction
when the 2DES density is raised above a critical density $\sim
2.9\times 10^{11}$ cm$^{-2}$ \cite{Zhu.PRL.2002}. The density-induced
rotation occurs at very similar densities for 2DESs confined to either
GaAs/AlGaAs hetero-junctions or GaAs quantum wells; also it does not
depend on the LL spin orientation as it happens for both filling
factors $\nu=9/2$ and 11/2 \cite{Zhu.PRL.2002,Cooper.PRL.2004}.



Here we report a study of stripe phases in wide GaAs quantum wells
(QWs) where two electric subbands are occupied. Our main focus is on
the evolution of the orientation of stripe phases as we increase the
density while keeping the QW charge distribution symmetric
(balanced). More precisely, we monitor the magneto-resistance near LL
filling factors $\nu=13/2$ and 15/2 when $E_F$ lies in the two,
spin-split, $N=2$ LLs of the symmetric (S) subband (the S2$\uparrow$
and S2$\downarrow$ levels) while the $N=0$ LLs of the antisymmetric
subband (A0$\uparrow$ and A0$\downarrow$ levels) are fully occupied.
We find that when $E_F$ lies in S2$\downarrow$ the stripes are always
formed along the "normal" ([110]) direction. But, when $E_F$ lies in
the S2$\uparrow$ level, the orientation of the stripes can rotate to
be along the "abnormal" ([1\=10]) direction at high densities. At a
density where the stripe phase at $\nu=13/2$ is along the abnormal
direction, we can rotate it back to the normal direction by making the
QW charge distribution asymmetric while keeping the density fixed. Our observations therefore reveal
that the symmetry-breaking mechanism that determines the direction of
the stripe phases depends not only on the 2DES density but also on the
$spin$ orientation of the LL in which $E_F$ resides, and on the
symmetry of the charge distribution in the QW.

\begin{figure*}[t]
\includegraphics[width=0.98\textwidth]{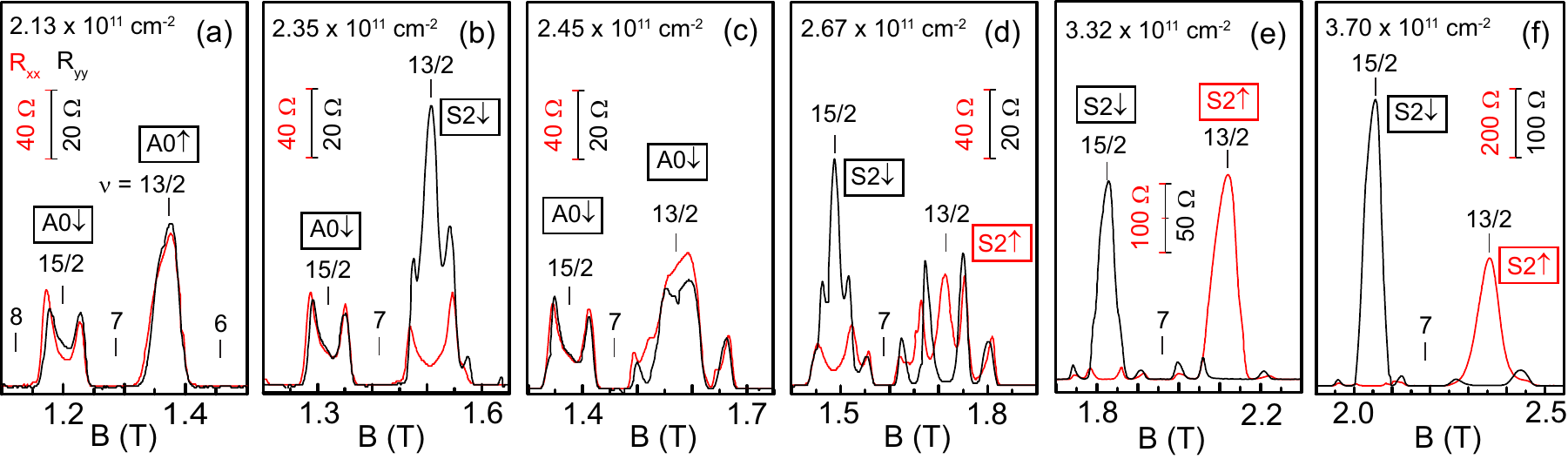}
\caption{\label{fig:42nm}(color online) Longitudinal
  magneto-resistance measured at $T=30$ mK in a 42-nm-wide QW along
  the [110] (${R_{xx}}$, red traces) and [1\=10] ($R_{yy}$, black
  traces) directions. Data are shown in the filling factor range
  $6<\nu<8$ at different electron densities as indicated. The field
  positions of half-filled LLs ($\nu=13/2$ and 15/2) are marked by
  vertical lines, and the LL in which $E_F$ resides at the different
  half-fillings are indicated in boxes.}
\end{figure*}

\begin{figure}[b]
\includegraphics[width=0.4\textwidth]{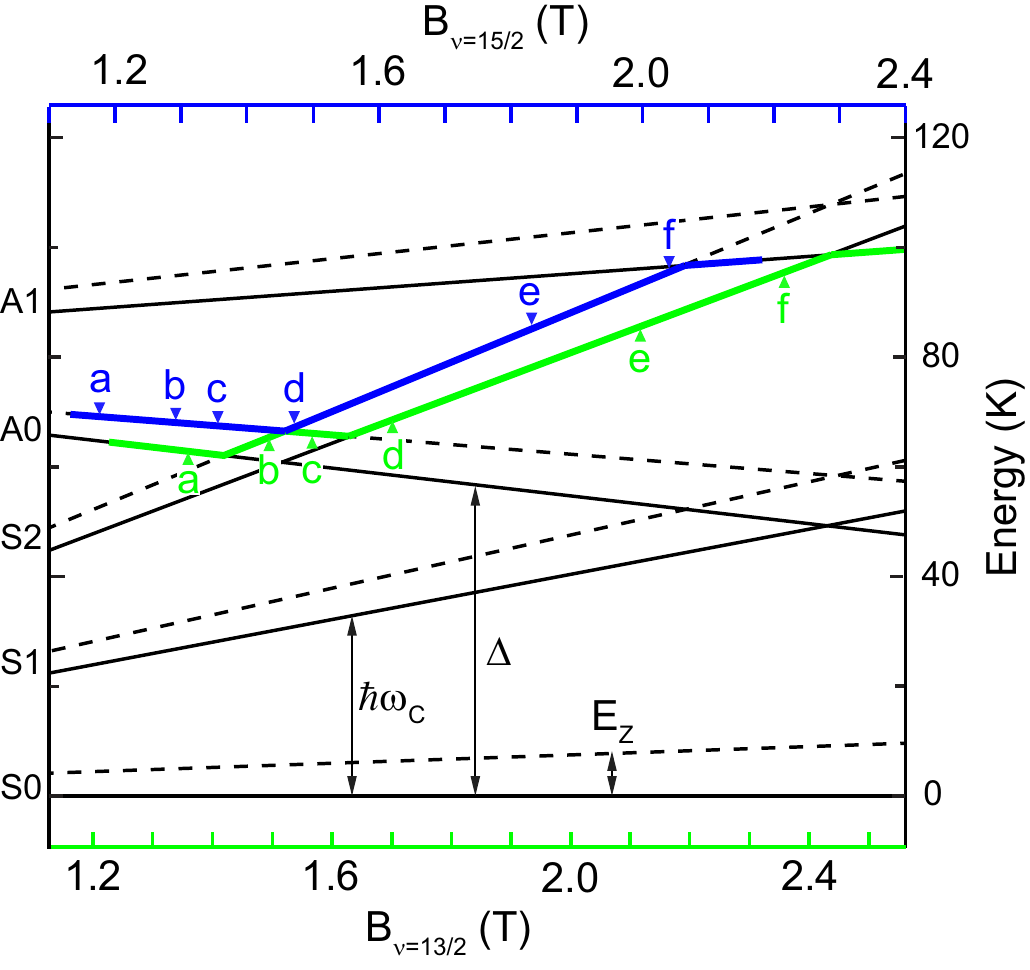}
\caption{\label{fig:LLdiag}(color online) Landau level energy diagram
  as a function of density for the 42-nm-wide QW. The relevant
  energies are the subband separation ($\Delta$), and the cyclotron
  and Zeeman energies ($\hbar\omega_c$ and $E_Z$); the up-
  ($\uparrow$) and down-spin ($\downarrow$) levels are represented by
  solid and dashed lines, respectively. The energies are plotted as a function of
  the field position of filling factor $\nu=13/2$ (bottom axis) and
  $\nu=15/2$ (top axis). The positions of $E_F$ at
  $\nu=13/2$ and 15/2 are marked by the green and blue lines,
  respectively. The triangles labeled a to f point to the positions of
  $\nu=13/2$ and 15/2 for the densities at which traces are shown in
  Figs. 1(a-f).}
\end{figure}

\begin{figure*}[t]
\includegraphics[width=0.98\textwidth]{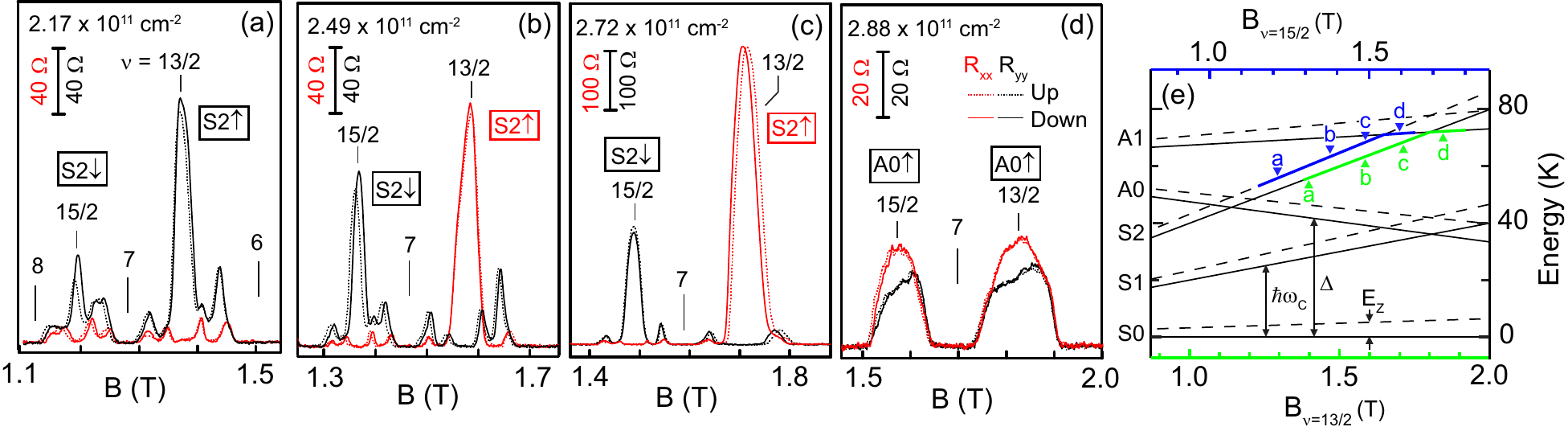}
\caption{\label{fig:51nm}(color online) (a-d) ${R_{xx}}$ and $R_{yy}$
  measured in a 51-nm-wide QW at four different densities as
  indicated. Solid (dotted) traces were taken while the field was
  swept down (up). All traces were taken at a sweep rate of 1 T/hour. (e) The LL energy fan diagram corresponding to the data shown
  in (a-d); the notations used are the same as those in Fig. 2.}
\end{figure*}

\begin{figure}[b]
\includegraphics[width=0.48\textwidth]{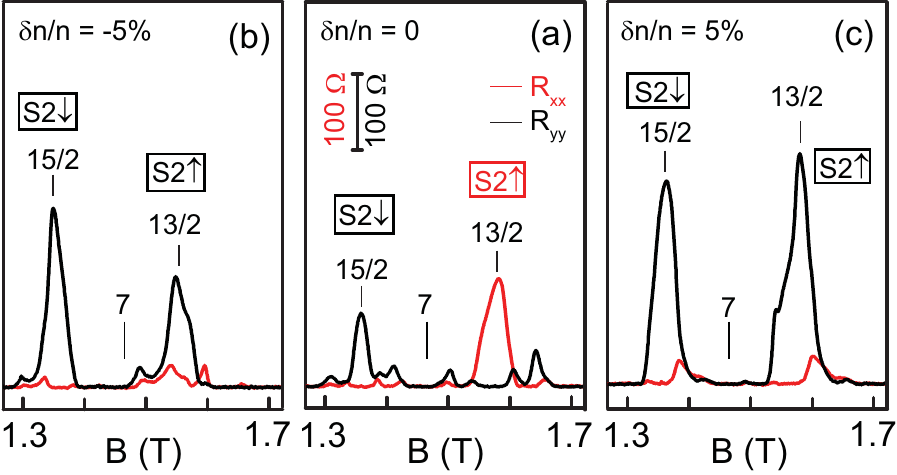}
\caption{\label{fig:51nmimb}(color online) (a) $R_{xx}$ and $R_{yy}$
  taken in the 51-nm-wide QW at a fixed density of $n=2.5\times
  10^{11}$ cm${^{-2}}$ when the QW charge distribution is
  symmetric. (b-c) Traces are shown as the charge distribution in the
  QW is made asymmetric by transferring $\sim5\%$ of total charge from
  the back side of the QW to the front side (b) or vice-versa (c).}
\end{figure}

Our samples were grown by molecular beam epitaxy, and each consist of
a wide GaAs QW bounded on either side by undoped
Al$_{0.24}$Ga$_{0.76}$As spacer layers and Si $\delta$-doped
layers. We report here data for two samples, with QW widths $W=$ 42
and 51 nm, and as-grown densities of $n\simeq$ 3.1 and 2.5 $\times
10^{11}$ cm$^{-2}$, respectively. The low-temperature ($T = 0.3$ K)
mobilities of these samples are $\mu \simeq$ 600 m$^2$/Vs. The samples
have a van der Pauw geometry and each is fitted with an evaporated
Ti/Au front-gate and an In back-gate. We carefully control the density
and the charge distribution symmetry in the QW by applying voltage
biases to these gates \cite{Suen.PRL.1994, Liu.PRB.2011}. The
measurements were carried out in a dilution refrigerator with base
temperature $T\simeq$ 30 mK, and we used low-frequency ($<20$ Hz)
lock-in techniques to measure the transport coefficients. Throughout
this article, the longitudinal resistances measured along the [110]
direction ($R_{xx}$) are shown in red, and those measured along the
[1\=10] direction ($R_{yy}$) are shown in black.  With this notation,
a black trace showing a much larger resistance than a red trace
corresponds to the "normal" stripe orientation ([110]), i.e., the one
that is commonly seen in standard, single-subband QWs at low
densities. Conversely, a black trace showing a much smaller resistance
than a red trace signals that the stripes are formed along [1\=10],
which we refer to as the "abnormal" orientation.

Figure 1 illustrates one of our main findings. It shows $R_{xx}$ and
$R_{yy}$ traces, in the filling range $6<\nu<8$, for a symmetric
(balanced) 42-nm-wide QW at six different densities $n=2.13$, 2.35,
2.45, 2.67 3.32, and $3.70\times 10^{11}$ cm$^{-2}$. At $\nu=13/2$, as
a function of increasing $n$, transport is first isotropic
(Fig. 1(a)), shows a "normal" anisotropy (Fig. 1(b)), becomes
isotropic again (Fig. 1(c)), and then exhibits anisotropy but now
along the "abnormal" direction (Figs. 1(d-f)). The behavior at
$\nu=15/2$, however, is markedly different; it is isotropic in
Figs. 1(a-c) and then shows a "normal" anisotropy at higher $n$ (Figs. 1(d-f)). The
traces shown in Figs. 1(d-f) are particularly noteworthy: transport is
anisotropic at both $\nu=13/2$ and 15/2, but the orientation of the
anisotropy in a $single$ $trace$ is different at these two fillings.

In order to understand the data of Fig. 1, we present in Fig. 2 a
Landau level (LL) fan diagram for this 42-nm-wide QW sample as a
function of $n$, or equivalently the magnetic field position of
$\nu=13/2$ ($B_{\nu=13/2}$).  We show the LLs for the symmetric (S)
and antisymmetric (A) electric subbands. The index 0, 1, or 2
following S and A is the LL orbital quantum number ($N$), and the up-
($\uparrow$) and down-spin ($\downarrow$) levels are represented by
solid and dashed lines. The relevant energies are the subband
separation ($\Delta$), the cyclotron energy ($\hbar\omega_c$), and the
Zeeman energy ($E_Z$). As we increase $n$ while keeping the QW
balanced, $\hbar\omega_c$ and $E_Z$ increase but $\Delta$ decreases
\cite{Suen.PRL.1994, Shayegan.SST.1996}, causing crossings of the S2
and A0 levels. As we discuss below, these crossings are consistent
with the evolution seen in Fig. 1. We emphasize
that the LL fan diagram shown in Fig. 2 is based $quantitatively$ on
the parameters of our sample. For example, we measured $\Delta$ from
Fourier transforms of the Shubnikov-de Haas oscillations at low
magnetic fields \cite{Suen.PRL.1994}. These measured $\Delta$ are also
consistent with all the parallel-spin LL crossings we observe in this
sample \cite{Liu.PRB.2011}; these crossings occur at
$\Delta=i\cdot\hbar\omega_c$ where $i=1,2,3...$. We found that the
expression $\Delta=80-8.2\cdot n$, which we use in Fig. 2 plot,
accurately describes the dependence of $\Delta$ on $n$ in the range of
densities reached in our experiments ($\Delta$ has units of K and $n$
is given in units of $10^{11}$ cm$^{-2}$). For $E_Z$ we used an
effective g-factor of $g^*=3.5$ which is 8-fold enhanced relative to
the GaAs band g-factor (0.44); this $E_Z$ is consistent with all the
observed crossings between LLs of antiparallel-spin, which are
signaled by spikes in the longitudinal resistance
\cite{Liu.PRL.2011B}.

Focusing first on $\nu=13/2$, in Fig. 2 we show the position of the
$E_F$ at this filling in green, and mark the densities (or
$B_{\nu=13/2}$) corresponding to the data of Fig. 1 with up-pointing
triangles. At the lowest density $n=2.13\times 10^{11}$ cm$^{-2}$,
$E_F$ lies in the A0$\uparrow$ level at $\nu=13/2$ and the 2DES is
isotropic as seen in Fig. \ref{fig:42nm}(a). As we increase $n$ to
$2.35\times 10^{11}$ cm$^{-2}$, $E_F$ moves to the S2$\downarrow$
level at $\nu=13/2$. Strong anisotropy is seen in the data
(Fig. 1(b)), consistent with $E_F$ now lying in an $N=2$ LL. The
resistance peak in $R_{yy}$ and minimum in $R_{xx}$ indicate the
stripe phase is along the "normal" direction. Further increasing $n$
to $2.45\times 10^{11}$ cm$^{-2}$, this stripe phase disappears and
the 2DES becomes isotropic again (Fig. \ref{fig:42nm}(c)) when $E_F$
moves back to an $N=0$ LL, namely the A0$\downarrow$ level. The
anisotropy reappears as soon as $E_F$ moves to the S2$\uparrow$ level
at $n=2.67\times 10^{11}$ cm$^{-2}$ (Fig. \ref{fig:42nm}(d)) and the
2DES remains anisotropic up to the highest $n$ achievable in this
sample. Remarkably, however, in Figs. 1(d-f) at $\nu=13/2$ we observe a resistance
$peak$ in $R_{xx}$ and a $minimum$ in $R_{yy}$, signaling that the
stripe direction has rotated and is now along the "abnormal"
direction.

At $\nu=15/2$, transport is isotropic at the lowest three $n$
(Figs. 1(a-c)). This is expected as $E_F$ lies in the A0$\downarrow$
level; see the blue lines and the down-pointing triangles in
Fig. 2. When $n$ is further increased, $E_F$ moves to the
S2$\downarrow$ level and the 2DES becomes anisotropic (Figs. 1(d-f))
at $\nu=15/2$. In sharp contrast to the $\nu=13/2$ case, however, the
stripe phase at $\nu=15/2$ is oriented along the "normal" direction up
to the highest $n$ achievable in the sample. It is clear that at a
given fixed density (e.g., Fig. 1(e)), the stripes' direction depends
on the spin orientation of the LL where $E_F$ resides (S2$\uparrow$
for $\nu=13/2$ and S2$\downarrow$ for $\nu=15/2$).

Data taken in a 51-nm-wide QW (Fig. \ref{fig:51nm}) qualitatively
confirm the spin-dependent reorientation of the stripe phase. As we
increase $n$, the stripe phase rotates from the normal to the abnormal
direction if $E_F$ lies in the S2$\uparrow$ level at $\nu=13/2$, but
it never rotates when $E_F$ lies in the S2$\downarrow$ level at
$\nu=15/2$ (Figs. 3(b,c)). However, the reorientation at $\nu=13/2$ is
not seen at the lowest $n$ (Fig. 3(a)), suggesting that it depends on
$n$ also. Figure 3(d) indicates that, as expected, the 2DES becomes
isotropic at the highest $n=2.9\times 10^{11}$ cm$^{-2}$ when $E_F$
moves to the A1 LLs (see Fig. 3(e)). Note also that in Figs. 3(a-d) we
are showing data for different magnetic field sweep directions. In
contrast to previous observations near the stripe phase reorientations
in single-subband 2DESs \cite{Zhu.PRL.2002,Cooper.PRL.2004}, we
observe no hysteresis in our data \footnote{To be more precise, our bidirectional field sweep
data on the 51-nm-wide sample were taken at densities which are about 5\% away from the transition density. We cannot rule out that hysteresis might exist closer to the transition density.}.

Figure 4 illustrates yet another remarkable property of the stripe phases
in our samples. Here data are shown for the 51-nm-wide sample of
Fig. 3 at a fixed density of $n=2.5\times 10^{11}$ cm$^{-2}$ while we
make the charge distribution in the QW asymmetric via applying front-
and back-gate voltage biases with opposite polarity. When the charge
distribution is symmetric (Fig. 4(a)) the stripe phase at $\nu=13/2$
is along the abnormal direction, but a small asymmetry in the charge
distribution reorients the phase along the normal
direction \footnote{At a higher density of $n=2.7\times10^{11}$
  cm$^{-2}$, the stripe phase at $\nu=13/2$ in the 51-nm-wide QW sample remains along the
  abnormal direction when $\delta n/n\simeq$ 4\%.}.

The data presented in Figs. 1-4 provide evidence for additional
subtleties and twists in the physics of stripe phases in 2DESs. While
we do not have an explanation for the behaviors revealed in our wide
QW data, some implications are noteworthy. First, in both the 42- and
51-nm-wide QWs, the reorientation at $\nu=13/2$ occurs at a
very similar density, $n\simeq2.5\times 10^{11}$ cm$^{-2}$
\footnote{Also, in a 65-nm-wide QW, we do not see any rotation of the
  stripe phase at $\nu=13/2$ up to the highest achievable density of
  $n=2.0\times 10^{11}$ cm$^{-2}$.}. Therefore, we cannot rule out the
possibility that our observed reorientation is
density-induced. However, in single-subband, narrow QWs, the stripe
phases at $\nu=9/2$ and 11/2 both rotate above the same threshold
density ($\sim 2.9\times 10^{11}$ cm$^{-2}$, see
\cite{Zhu.PRL.2002}), suggesting that the electron spin is not playing
a role. In contrast, the rotation we report here in wide QWs appears
to be spin-dependent: the stripe phase rotates at $\nu=13/2$ when
$E_F$ lies in the S2$\uparrow$ level, but never rotates at $\nu=15/2$
when $E_F$ is in the S2$\downarrow$ level. Also, in our samples the
stripe phase rotates at a density ($n\simeq 2.5 \times 10^{11}$
cm$^{-2}$) which is smaller than the well-established critical density
$n\simeq 2.9 \times 10^{11}$ cm$^{-2}$ in hetero-junctions and narrow
QW samples \cite{Zhu.PRL.2002, Cooper.PRL.2004}. Moreover, the filling
factors in our study ($\nu=13/2$ and 15/2) are larger than in previous
reports (9/2 and 11/2). Together with the lower threshold densities,
this implies that the transition fields in our experiment are much
smaller compared to previous measurements ($\sim$ 1.5 T vs. $\sim$ 2.8
T).

Second, as illustrated in Fig. 4, the rotated stripe phase can be
switched back to the normal direction when the QW charge distribution
is made asymmetric. This observation has important
implications for the possible origins of the symmetry-breaking potential. For example, Koduvayur \emph{et al.}
\cite{Koduvayur.PRL.2011} recently reported that the application of in-plane
shear strain can alter the exchange potential and re-align the stripe
direction in GaAs 2D hole systems. Thus they suggested that the
residual strain due to surface charge induced fields is responsible
for the symmetry-breaking potential of the stripe phases in both hole
and electron 2D systems in GaAs. Our data of Fig. 4 do not agree with
this conjecture as they show that the stripe phase can be made to lie
along the same (normal) direction for electric fields of opposite
polarity.


The experimental observations reported here point to additional
intricacies that determine how a GaAs 2DES chooses the direction of
its anisotropic (stripe) phases at half-filled LLs. Besides the 2DES
density, the spin orientation of the LL where $E_F$ lies, as well as
the symmetry of the charge distribution can both play roles in
stabilizing the stripe phase direction. The spin-dependence is
particularly puzzling because the energy of a stripe phase normally
should not depend on the spin orientation of the carriers. It is
possible that factors such as the mixing of the nearby LLs,
particularly the A1 LLs, are responsible for the spin-dependence we
observe. The details we preset here, namely, our samples' parameters
(well width, density, charge-distribution symmetry, and LL energy
diagrams) should provide stimulus and quantitative input for future work aimed at
understanding what determines the orientations of the stripe phases.

\begin{acknowledgments}
  We acknowledge support through the NSF (grants DMR-0904117 and MRSEC
  DMR-0819860), and the Moore and Keck Foundations. This work was
  performed at the National High Magnetic Field Laboratory, which is
  supported by the NSF Cooperative Agreement No. DMR-0654118, by the
  State of Florida, and by the DOE. We thank E. Palm, J. H. Park,
  T. P. Murphy and G. E. Jones for technical assistance.
\end{acknowledgments}

\bibliographystyle{apsrev4-1}
\bibliography{../bib_full}

\end{document}